# Photo-Activated, Solid-State Introduction of Luminescent Oxygen Defects into Semiconducting Single-Walled Carbon Nanotubes

*Sonja Wieland, Abdurrahman Ali El Yumin, Simon Settele, Jana Zaumseil\**

Institute for Physical Chemistry, Universität Heidelberg, D-69120 Heidelberg, Germany

Corresponding Author

*E-mail: zaumseil@uni-heidelberg.de

**ABSTRACT**

Oxygen defects in semiconducting single-walled carbon nanotubes (SWCNTs) are localized disruptions in the carbon lattice by the formation of epoxy or ether groups, commonly through wet-chemical reactions. The associated modifications of the electronic structure can result in luminescent states with emission energies below those of pristine SWCNTs in the near-infrared, which makes them promising candidates for applications in bio-sensing and as single-photon emitters. Here, we demonstrate the controlled introduction of luminescent oxygen defects into networks of monochiral (6,5) SWCNTs using a solid-state, photocatalytic approach. UV irradiation of SWCNTs on the photoreactive surfaces of the transition metal oxides $TiO_x$ and $ZnO_x$ in the presence of trace amounts of water and oxygen results in the creation of reactive oxygen species that initiate radical reactions with the carbon lattice and the formation of oxygen defects. The created ether-d and epoxide-l defect configurations give rise to two distinct red-shifted emissive features. The chemical and dielectric properties of the photoactive oxides influence the final defect emission properties, with oxygen-functionalized SWCNTs on $TiO_x$ substrates being brighter than on $ZnO_x$ or pristine SWCNTs on glass. The photoinduced functionalization of nanotubes is further employed to create lateral patterns of oxygen defects in (6,5) SWCNT networks with micrometer resolution and thus spatially controlled defect emission.

**INTRODUCTION**

The distinct optical characteristics of semiconducting single-walled carbon nanotubes (SWCNTs) arise from their quasi-one-dimensional structure and the formation of tightly bound excitons upon their excitation.[1-2] The strongest absorption and emission peaks are a manifestation of the bright $E_{11}$ exciton. They are observed at diameter-specific energies in the near-infrared range (>900 nm).[3] Of the 16 possible excitons in SWCNTs only one is bright, with most of the dark excitons lying energetically lower than the bright exciton.[4-5] These dark excitons are one cause of the generally low photoluminescence quantum yield (PLQY) of SWCNTs.[6] Another reason is the fast diffusion of excitons to structural defects or nanotube ends, where they can decay non-radiatively.[7] However, certain defects in the carbon lattice do not act as quenching sites that reduce the PLQY but form lower-lying states that trap mobile excitons and enable radiative decay at longer wavelengths.[8-9] Such luminescent defects can be formed by covalent functionalization, for example by aryl diazonium chemistry, to create sp³ carbon atoms within the sp² carbon lattice.[10] The resulting defect emission energies ($E_{11}$* or $E_{11}$*⁻) are mainly determined by the binding configuration of the sp³ carbons,[11] which can be controlled through the choice of synthetic conditions.[12-14] Very similar emission features and indeed substantially brighter photoluminescence (PL) are obtained through the reaction of nanotubes with reactive oxygen species.[15] The properties of the resulting luminescent oxygen defects also depend on the precise binding configuration, with the oxygen being integrated in an ether or epoxide structure (see **Figure 1a**).[16] Note that the hybridization state of the neighboring C atoms (sp² vs. sp³) further influences the defect properties, as observed previously for oxygen- and nitrogen-functionalized SWCNTs.[17-19] Oxygen defects are typically introduced through wet-chemical approaches such as reactions of ozone or hypochlorite in water under UV illumination,[15-16, 20-21] but have also been observed after electron beam evaporation of metal oxides onto SWCNT networks.[22] SWCNTs with luminescent defects can be applied for in-vivo imaging in the second biological window,[23-24] optical sensing of biomarkers,[25] and single-photon emission at room temperature.[26]

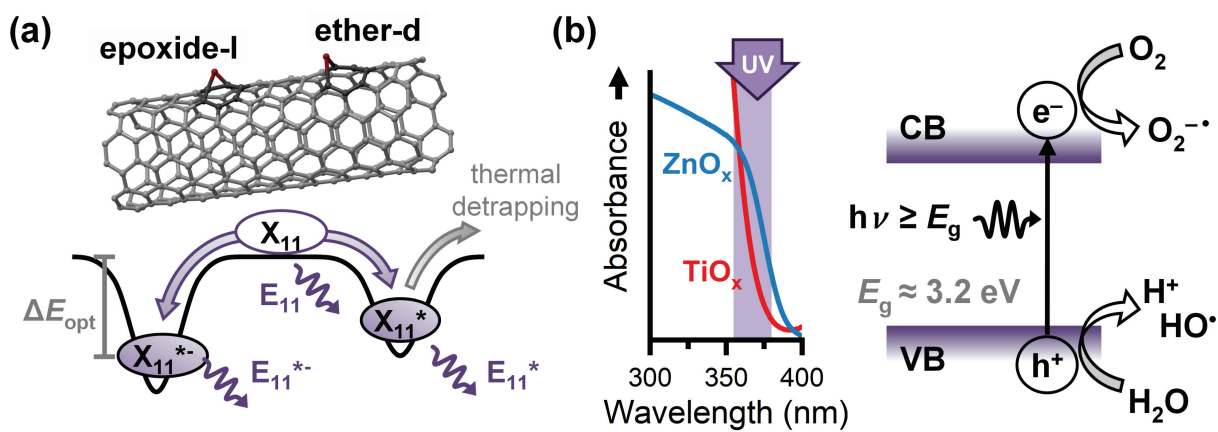

**Figure 1.** a) Possible oxygen defect configurations and energetics of different mobile ($X_{11}$) and localized excitons ($X_{11}^{*-}$ and $X_{11}^{*}$) in SWCNTs, including the corresponding emission energies ($E_{11}$, $E_{11}^{*-}$ and $E_{11}^{*}$) and optical trap depth $\Delta E_{opt}$ (energy difference between the mobile exciton and defect emission). b) Absorption edge spectra of $TiO_x$ and $ZnO_x$ (UV excitation wavelength range used here is indicated in purple) and schematic of photo-induced creation of reactive oxygen species (ROS) by oxidation of water and reduction of oxygen at the surface. CB – conduction band, VB – valence band, $E_g$ – band gap energy.

The creation of luminescent oxygen defects requires less experimental effort than $sp^3$ defects with aryl or alkyl groups. However, unintentional overfunctionalization and degradation of nanotubes during sample processing and storage under ambient conditions might become an issue. Although nanotubes are overall much more stable than typical organic fluorophores as evidenced by their photostability under laser excitation[27] and high thermal decomposition temperatures (>400 °C),[28] there is significant evidence that (photo)oxidation of SWCNTs can alter SWCNT properties in undesired ways.[29-30] Furthermore, Zorn *et al*. observed that SWCNTs strongly interact with polar silicon dioxide or glass surfaces, which are frequently used as dielectrics or substrates in optoelectronic devices, leading to defect-related broadening and emission sidebands especially after high-temperature annealing.[29] However, these undesired effects could be avoided by substrate passivation with hexagonal boron nitride (*h*-BN) or crosslinked non-polar polymers (e.g., BCB). In contrast to that, more reactive or photocatalytic oxidic surfaces in contact with SWCNTs might have a much larger impact on the emission properties of SWCNTs. They may lead to even stronger photodegradation or could be used for the controlled introduction of luminescent defects in pre-deposited nanotubes.

(6,5) SWCNTs are widely employed for fundamental studies of luminescent defects[31] as they can be sorted in substantial quantities and with high chiral purity through various methods such as selective polymer-wrapping in organic solvents,[6] aqueous two-phase extraction[32] and gel-chromatography[33] for aqueous dispersions. The high strain in these small-diameter nanotubes increases their reactivity[34] and facilitates defect introduction.[8] Their relatively large band gap around 1.24 eV enables simple detection even of further red-shifted defect emission features.

The transition metal oxide semiconductors $TiO_2$ and ZnO are well-known as heterogeneous photocatalysts for the degradation of organic pollutants and the production of renewable fuels and hydrogen by water splitting.[35-36] Excitation with energies greater than their band gap (~ 3.2 eV), that is, UV irradiation, creates electron-hole pairs that can participate in surface redox reactions, primarily with oxygen and water (see **Figure 1b**). Under ambient conditions, oxygen is reduced to the reactive superoxide radical anion ($O_2^{-\bullet}$), while water is oxidized to the even more reactive hydroxy radical (HO•).[37-39] These reactive oxygen species (ROS) readily initiate radical reactions with other molecules and specifically carbon-based materials directly on the surface.

Here, we demonstrate the photoinduced introduction of luminescent oxygen defects in sparse networks of polymer-sorted (6,5) SWCNTs on TiO$_2$ and ZnO surfaces upon UV-light exposure in the presence of only trace amounts of water and oxygen. This solid-state photoreaction enables patterning of luminescent defects in nanotube networks with micrometer resolution. Unlike previous approaches to photopatterning of defects,[40-41] this method is chemical- and solvent-free.

## METHODS

**Preparation of (6,5) SWCNT Dispersions**

Shear-force mixing (Silverson L2/Air, 10230 rpm) of 50 mg CoMoCat raw material (Sigma-Aldrich, MKCJ7287) in a solution of 65 mg of PFO-BPy (poly[(9,9-dioctylfluorenyl-2,7-diyl)-*alt*-(6,6'-(2,2'-bipyridine))], American Dye Source, Inc, $M_w$ = 40 kg mol$^{-1}$) in 140 mL of toluene for 72 h at 20 °C was used to prepare nearly monochiral (6,5) SWCNT dispersions as described previously.[6] Non-dispersed amorphous carbon and other SWCNT chiralities where separated from the PFO-BPy-wrapped (6,5) SWCNTs by centrifugation (twice, 45 min at 60000 *g*, Beckman Coulter Avanti J26XP centrifuge) and filtration through a syringe filter (Whatman PTFE membrane, pore size 5 µm). To remove excess PFO-BPy, the polymer-rich stock dispersion was vacuum-filtered through a polytetrafluoroethylene membrane (Merck Millipore, JVWP, pore size 0.1 µm). The obtained filter cake was washed three times in 10 mL of toluene at 80 °C for 10 min and redispersed in fresh toluene by bath sonication for 30 min. Absorption (Cary 6000i UV-vis-NIR spectrometer, Varian Inc.) and Raman spectroscopy (InVia Reflex, Renishaw plc) confirmed the chiral purity and low content of unbound polymer of the (6,5) SWCNT dispersion (see **Figure S1, Supporting Information**).

**Substrate preparation**

Different reactive and unreactive surface layers were prepared on cleaned glass slides (sodium-free aluminum borosilicate glass, Schott AF32 eco). 9 nm thick titanium dioxide layers were prepared by annealing a spincoated organotitanate precursor layer (Solaronix Ti-Nanoxide BL/SC) at 500 °C for 45 min. Zinc oxide layers were deposited by airbrush-spraying of a 0.3 molar solution of $Zn(OAc)_2 \cdot 2\,H_2O$ (Sigma Aldrich, 99.9 %) in methanol onto a heated (200 °C) glass substrate, resulting in 200 nm thick layers. For a passivated surface without any oxidic or hydroxyl groups, divinyltetramethyl-siloxanebisbenzocyclobutene (BCB) precursor (Cyclotene 3022–35, micro resist technology GmbH) was diluted 1:4 in mesitylene, spin-coated and subsequently cross-linked at 290 °C for 2 min, resulting in 80 nm layers.

All substrates (except BCB-passivated substrates) were treated in a UV/ozone cleaner for 10 min before depositing the (6,5) SWCNTs from toluene dispersion ($E_{11}$ absorbance of 2 at 1 cm path length) with three consecutive spin-coating steps. Samples were heated at 90 °C for 2 min after each step. Excess PFO-BPy was removed by washing with tetrahydrofuran and 2-propanol and subsequent annealing at 90 °C for 4 min. The deposited (6,5) SWCNTs had an average length of $1.2 \pm 0.4$ µm and showed network densities of around 10 nanotubes/µm, as determined by atomic force microscopy (AFM, Bruker Dimension Icon, see **Figure S2, Supporting Information**).

**Defect introduction**

The (6,5) SWCNT networks on different surfaces were illuminated with a UV light-emitting diode (SOLIS-365C, Thorlabs, 365 nm, 1.9 mW mm$^{-2}$) for 2, 15 and 30 min in ambient air or in dry nitrogen (glovebox, $H_2O$: 0.7 ppm, $O_2$: < 0.1 ppm), or for 15 min with a green light-emitting diode (SOLIS-525C, Thorlabs, 525 nm, 1.4 mW mm$^{-2}$, together with a 500 nm long-pass filter to exclude unintentional substrate excitation) in dry nitrogen. A fresh substrate was used for each irradiation condition. For the UV exposure series (exposure time-dependent experiments in air, in dry nitrogen), individual substrate pieces were cut from one parent substrate after spin-coating to ensure similar SWCNT densities across the exposure series. All substrates were annealed at 150 °C for 30 min inside the glovebox before illumination. Photo-patterning of defects was achieved by exposure with UV light through a 60 μm thick metal shadow mask with holes of 60 μm diameter for 30 min in dry nitrogen.

**Optical Characterization**

Photoluminescence (PL) spectra were collected with a Princeton Instruments IsoPlane SCT 320 spectrometer in combination with a thermo-electrically cooled InGaAs camera (NIRvana 640ST, Princeton Instruments, 512 × 640 pixels). An 850 nm long-pass filter in front of the spectrometer entrance slit blocked scattered excitation laser light (640 nm, continuous-wave laser diode, OBIS, Coherent). All spectra were averaged over an area of 800 μm$^2$ through expansion of the laser beam using a plano convex lens (focal length $f$ = 125 mm) in front of the near-infrared-optimized 20× Olympus objective and vertical binning to remove spot-to-spot variations as reported elsewhere.[42] Temperature-dependent PL measurements were carried out in a closed-cycle liquid helium cooled cryostat (Montana Cryostation s50) at high vacuum (< 10$^{-5}$ mbar) using a long-working distance, NIR-optimized 50× objective (N.A. 0.42, Mitutoyo) mounted outside the cryostat. PL spectra were corrected for the featureless scattering background of the excitation laser light and for absorption of optics in the detection path and sensitivity of the detector. They were fitted with three Lorentzian peaks after Jacobian wavelength-to-energy-scale conversion.[43]

Time-correlated single-photon counting (TCSPC) was carried out to determine PL lifetimes using an InGaAs/InP avalanche photodiode (Micro Photon Devices). The SWCNTs were excited at 640 nm with a picosecond-pulsed supercontinuum laser (NKT Photonics SuperK Extreme) and the detection wavelength was selected using a spectrograph (Princeton Instruments Acton SpectraPro SP2358). TCSPC time traces were generated with a counting module (PicoQuant PicoHarp 300) and fitted as biexponential decays with a short and a long lifetime component.

Raman spectra (Renishaw inVia confocal Raman microscope) were collected from 1500 spots across 40×70 µm² of the nanotube networks using an Olympus 50× short working distance objective under near-resonant excitation at 532 nm. Raman peaks were fitted for each spot with the Renishaw Wire 3.4 software using a combination of Lorentzian and Gaussian fit functions. Average values of the $D/G^+$ ratios were obtained by averaging the ratios obtained from single fits. However, for D-band intensities below 100 counts, i.e., in case of the $TiO_x$ sample irradiated for 2 min in air, the signal was not sufficient for reliable single-spot fitting and the $D/G^+$ ratios were calculated from fits of the averaged Raman spectra instead.

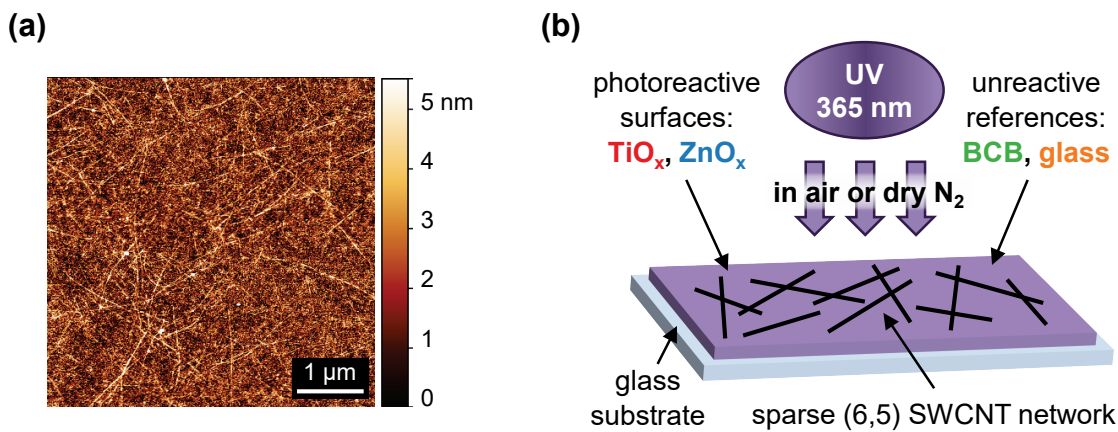

**Figure 2.** a) Atomic force micrograph of a sparse SWCNT network on $TiO_x$. b) Schematic substrate layout and UV exposure geometry: Sparse, spin-coated (6,5) SWCNT networks on a glass substrate coated with different photo-active ($TiO_x$, $ZnO_x$) and passivated or unreactive reference (BCB, glass) surfaces.

**RESULTS & DISCUSSION**

To explore the photo-activated introduction of luminescent oxygen defects into (6,5) SWCNTs on reactive surfaces, semiconducting transition metal oxides that are known for their photocatalytic behavior, such as $TiO_2$ and ZnO, were chosen as outlined above. The corresponding oxide layers were formed via spincoating or spray-coating on glass followed by annealing (see **Methods Section**). Nevertheless, they remained mostly amorphous and the precise stoichiometry may vary, hence, we refer to them as $TiO_x$ and $ZnO_x$ in the following. Clean aluminum borosilicate glass or BCB-passivated glass substrates served as reference samples. Deposition of (6,5) nanotubes by spin-coating from one stock dispersion onto different substrates gave sparse and random networks with similar densities on all surfaces (see **Figure 2a** and **Figure S2, Supporting Information**). After annealing in nitrogen, these networks were exposed to UV light (365 nm, 3.4 eV) in air or in dry nitrogen for different periods of time as schematically shown in **Figure 2b**. We can assume that all SWCNTs within the sparse network are in direct contact with the underlying substrate and, in contrast to dense networks or SWCNT bundles, are equally likely to undergo reactions with reactive species that are created at the oxide surface. Sparse networks further promote effective UV absorption by the underlying reactive oxide as light absorption by SWCNT/PFO-BPy will be negligible.

Before illumination, all as-deposited SWCNT networks (see **Figure 3a**) exhibited PL spectra with the characteristically narrow excitonic $E_{11}$ emission at 1.232 eV and weak, red-shifted sideband features, as commonly observed in networks of (6,5) SWCNTs. The sideband contributions to the PL signal increased from non-polar BCB to polar substrates and from nonreactive to reactive oxide surfaces as expected.[29, 44] These sidebands have been assigned to momentum-forbidden dark excitons coupling to phonons and to shallow extrinsic defects of the (6,5) SWCNT lattice due to interaction with reactive groups on mostly polar surfaces.[45-47] The additional feature observed at 1.051 eV for the TiO$_x$ sample most likely corresponds to SWCNT trion (charged exciton) emission due to slight p-doping by the oxide.[42] The relatively strong sideband for (6,5) SWCNTs even on BCB-passivated samples probably results from the low network density. All nanotubes were in direct contact with the underlying substrate and their remaining surface was exposed to the environment. Note that even the surface of polymer-wrapped nanotubes is only partially covered by polymer (appr. 10%).[48] Consequently, these SWCNTs were more susceptible to any unwanted reactions than nanotubes in dense networks or bundles.

**Figure 3a** further shows the impact of mild annealing (30 min at 150 °C) of the samples in the dry nitrogen atmosphere of a glovebox. This annealing step was performed to remove adsorbed water from the SWCNTs and the substrate surfaces as much as possible and to ensure reproducible starting points for all samples prior to UV exposure. Importantly, this treatment did not significantly alter the emission properties or the Raman spectra (specifically the intensity of the defect-related D-mode and the corresponding D/G$^+$ ratios as a metric for defect density,[49-50] see **Figure S3, Supporting Information**) of the SWCNT networks. For nanotubes on the TiO$_x$ surface, the trion emission disappeared and a featureless sideband similar to that on glass emerged while the Raman D/G$^+$ ratio increased slightly. While interactions with the polar substrate surface could contribute to the PL spectral shape, slight changes might also be indicative of thermally driven defect introduction in SWCNTs on highly reactive TiO$_x$ surfaces.

When these SWCNT networks were exposed to UV light in air, that is, in the presence of plenty of oxygen and water, their PL was almost completely quenched within less than 30 minutes. For nanotube networks on the highly reactive TiO$_x$ surface, the PL vanished completely after only 2 minutes of UV illumination in air (see **Figure 3b** and **Figure S4a, Supporting Information**).

The emission from nanotubes on $ZnO_x$ was quenched slower than on $TiO_x$, but still much faster than for those on glass or BCB. The rapid increase of the Raman $D/G^+$ ratios up to 0.5 upon UV illumination in air (**Figure S4b** and **Figure S4c, Supporting Information**) indicates the immediate creation of structural defects in the carbon lattice. For nanotubes on both $TiO_x$ and $ZnO_x$, we can assume that reactive oxygen species (ROS) such as $O_2^{-\bullet}$ and $HO\bullet$ were formed from oxygen and water, respectively, upon photoexcitation of the oxide.[35-36] These ROS directly attacked the carbon lattice of the nanotubes and created a large number of defects that led to non-radiative decay and PL quenching. The direct generation of reactive oxygen species from air through UV excitation of oxygen as demonstrated by Xhyliu *et al.*[51] for irradiation with 254 nm (4.88 eV) light is unlikely since the energy (3.4 eV) of the 365 nm light used here is too low for this process, but well above the bandgap of $TiO_x$ and $ZnO_x$ (optical band gaps of 3.19 eV for $ZnO_x$ and 3.29 eV for $TiO_x$ as determined by Tauc plots[52]). The much slower photodegradation of nanotubes on BCB and glass further highlights the important photocatalytic role of the oxides. PL quenching after several minutes of illumination is, however, consistent with reports by Larson et al., who proposed singlet oxygen sensitization or the formation of superoxide from oxygen through the direct excitation of the nanotubes in the UV to visible range as a cause for photodegradation.[30]

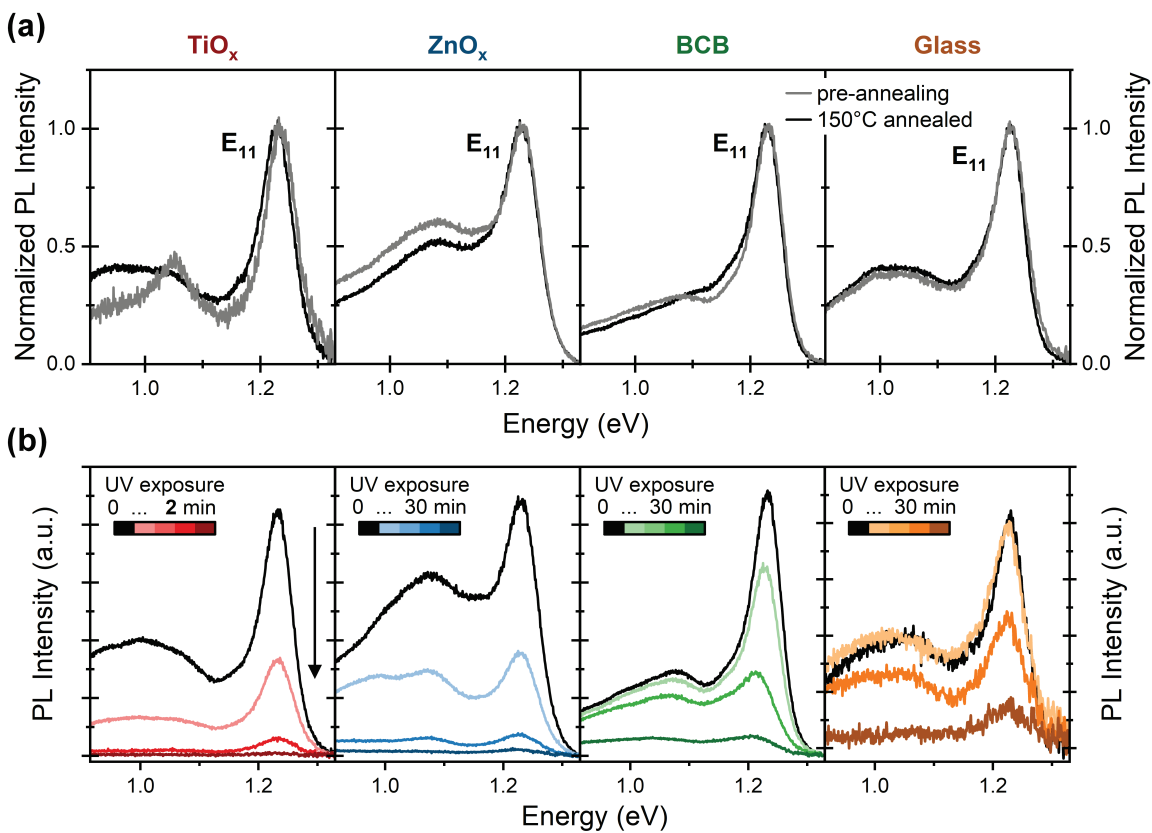

**Figure 3.** a) PL spectra of sparse (6,5) SWCNT networks on different substrates and effect of annealing at 150 °C for 30 min in dry nitrogen before illumination. b) Exposure time-dependent PL spectra of (6,5) SWCNT networks illuminated with UV-light in ambient air (TiO$_x$: 0, 5, 25, 120 sec, all others: 0, 2, 15, 30 min).

Importantly, at no point the spectral features of luminescent oxygen defects appear for nanotubes on any of the substrates. Either the sp² lattice is destroyed too quickly in too many places to allow for defect PL to be observed, or the type of defect formation is different under these conditions. ROS could initiate a chain reaction with excess oxygen, creating higher-oxidized species as previously observed for less-extended aromatic systems.[36] This should result in sidewall damage similar to the reaction products of superoxide treatment and oxidative cutting of SWCNTs, where e.g. hydroxyl, aldehyde, ketone and carboxylic acid defects act as exciton quenching sites and hence lead to a reduction of the PL intensity.[30, 53-54]

Clearly, the number of possible ROS produced by UV illumination must be reduced substantially to create a limited number of luminescent defects (only a few per micrometer) instead of damaging the entire carbon lattice. A much lower concentration of oxygen and water surrounding the nanotubes can be achieved by performing UV exposure inside a dry nitrogen glovebox. The previously described annealing step should also reduce adsorbed water or oxygen on the photoactive substrates, however, neither can be fully removed from a polar surface under these conditions (that is, not in high vacuum, limited temperature range). Residual ppm amounts of both oxygen and water will always be present.

**Figure 4a** shows the evolution of the PL spectra (normalized to $E_{11}$) of (6,5) SWCNT networks on different substrates after UV illumination for 2, 15 and 30 minutes in dry nitrogen. The PL spectra for nanotubes on BCB and glass did not change significantly in shape or intensity (see also **Figure S5, Supporting Information**). Their $D/G^+$ ratios also remained nearly constant (see **Figure 4b**). This indifference of SWCNTs on BCB and glass to UV exposure is in stark contrast to nanotubes on the photo-reactive oxide surfaces. After exposure of (6,5) nanotubes on both $TiO_x$ and $ZnO_x$, two additional red-shifted emission peaks, denoted here as $E_{11}^*$ (1.10 eV, 1130 nm) and $E_{11}^{*-}$ (0.95 - 0.97 eV, 1280 - 1300 nm) appeared (see **Figure S6, Supporting Information** for extended energy range spectra). The Raman $D/G^+$ ratios increased noticeably, but not to a degree that would indicate severe lattice damage.

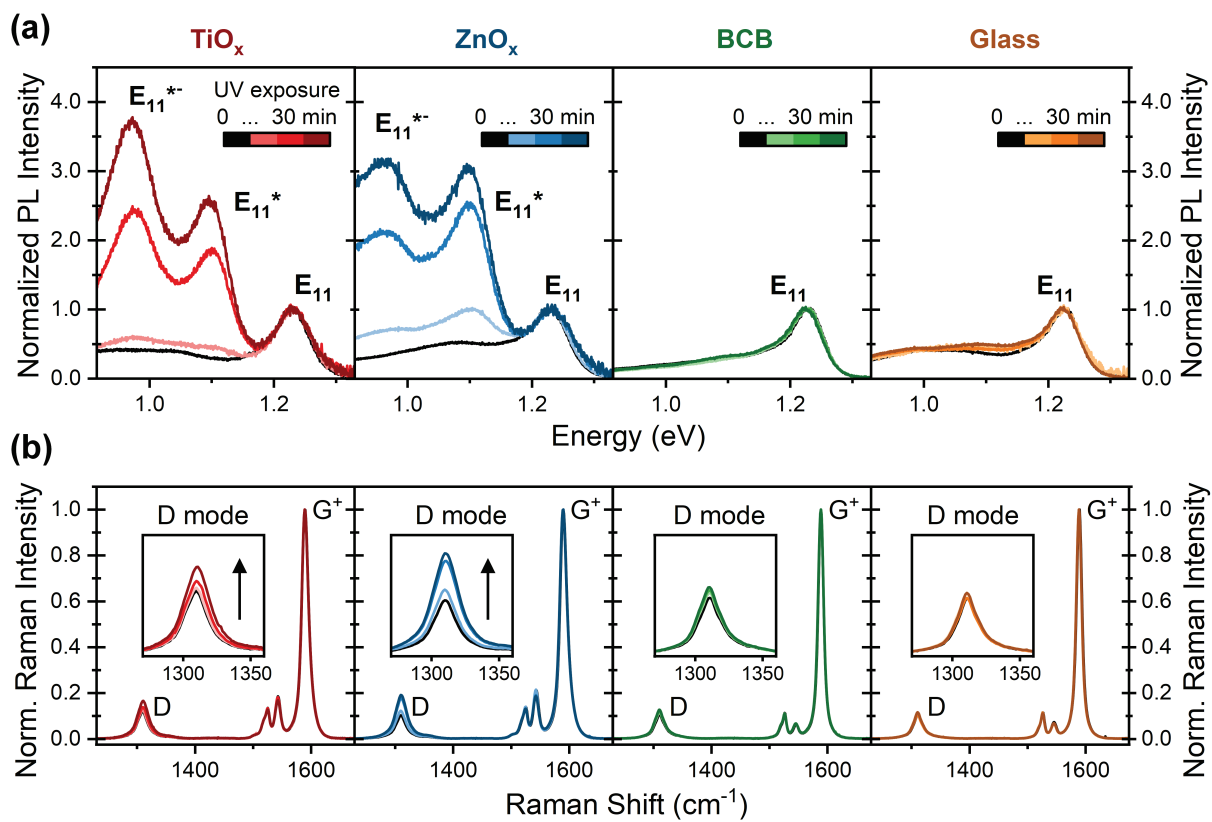

**Figure 4.** Evolution of PL spectra normalized to $E_{11}$ after UV exposure for 0, 2, 15, 30 min in dry nitrogen (glovebox). (b) Resonant Raman spectra (excitation at 532 nm) normalized to the $G^+$ mode show the increase of the defect-related D-mode with illumination for reactive surfaces.

The intensities of $E_{11}^*$ and $E_{11}^{*-}$ emission relative to the $E_{11}$ emission differed between the two photo-reactive oxides. UV-illuminated nanotubes on $ZnO_x$ exhibited broader defect emission (full width at half maximum (FWHM) $E_{11}^{*-}$: 214 meV, $E_{11}^*$: 96 meV) than SWCNTs on $TiO_x$ (FWHM $E_{11}^{*-}$: 163 meV, $E_{11}^*$: 86 meV), possibly caused by the higher dipolar disorder of the rough $ZnO_x$ surface.[44, 55] The overall PL intensity was actually enhanced for nanotube networks on $TiO_x$, but decreased for nanotubes on $ZnO_x$, and remained nearly constant for BCB substrates (see **Figure S5, Supporting Information**). The defect emission intensities relative to $E_{11}$ increased with reaction time (see **Figure 5a**) while the absolute $E_{11}$ intensities decreased due to trapping of mobile excitons at the defect sites. The almost linear increase of defects (see **Figure 5a**) with UV illumination time suggests that the reaction was limited by the number of formed ROS and no autocatalytic process or chain reaction was initiated. As shown in **Figure 5b**, the defect emission to $E_{11}$ ratio also increased linearly with the differential Raman $D/G^+$ ratio for nanotubes on both $TiO_x$ and $ZnO_x$ as observed before for luminescent $sp^3$ defects,[12, 56] with the slope being independent of the oxide type.

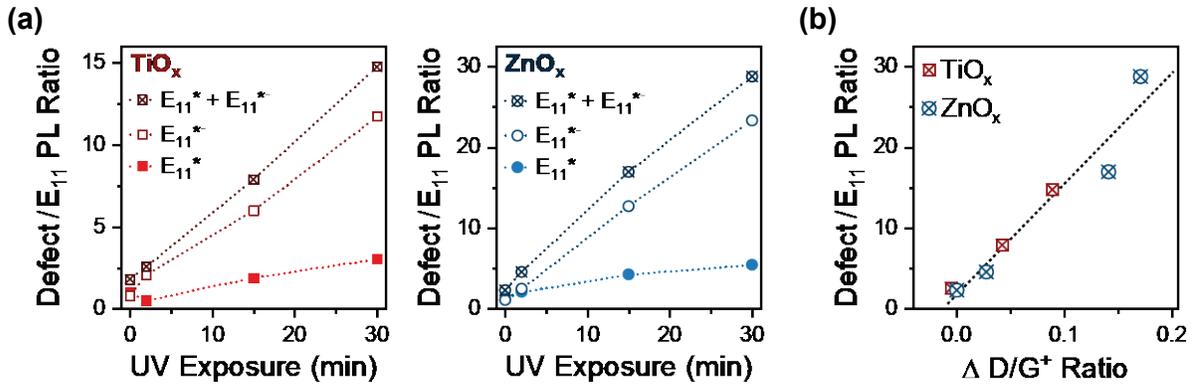

**Figure 5.** a) Increase in fitted relative defect PL area with UV exposure time in dry nitrogen for $TiO_x$ (left) and $ZnO_x$ (right). Note that the value at 0 min exposure time corresponds to intrinsic defect-related sideband emission of the pristine SWCNTs. b) Linear correlation of relative defect PL versus corresponding Raman $D/G^+$ ratio change induced by UV illumination of sparse SWCNT networks on photoactive surfaces ($TiO_x$, $ZnO_x$) in dry nitrogen.

The two main defect emission features $E_{11}^*$ and the more red-shifted $E_{11}^{*-}$ exhibit optical trap depths of 137 and 268 meV on $TiO_x$ and 140 and 288 meV on $ZnO_x$, respectively. These values are comparable to the trap depths of the ether-d (135 meV) and epoxide-l (300 meV) defect configurations observed by Ma. *et al.* after electron beam evaporation of oxides onto SWCNTs[16] and were also proposed for defects created by the reaction of nanotubes with ozone or hypochlorite under UV illumination.[15, 22-23] Hence, we assume that these or similar luminescent oxygen defects are formed by the reaction of ROS with the (6,5) nanotube lattice. The photoexcitation of $TiO_x$ and $ZnO_x$ by UV light should result in the formation of ROS in the vicinity of the nanotubes but in very low numbers due to the very limited concentrations of water or oxygen. The roles of the two different types of initially formed ROS, i.e., the superoxide radical anion and the hydroxy radical, cannot be distinguished directly. However, the formation of hydroxy radicals from water seems most likely as adsorbed water on polar surfaces is very difficult to remove. The hydroxy radical is also known to be more reactive.[37-39]

Previous reports have suggested that the bandgap excitation of nanotubes could result in electron transfer and photoinduced reactions to form luminescent defects without the need of high-energy UV light or the presence of photoreactive surfaces.[30, 57] As shown above, there is little or no change in the emission spectra of (6,5) SWCNTs when exposed to UV light on glass or BCB surfaces in dry nitrogen and only PL quenching over time when exposed to UV light in air. Nevertheless, we also tested whether luminescent defects could be introduced to SWCNTs on reactive surfaces when exposed to visible light that should not lead to any photoexcitation of $TiO_x$ or $ZnO_x$ but is above the bandgap and indeed excites the $E_{22}$ transition (575 nm, 2.16 eV) and phonon sideband of (6,5) nanotubes (see **Figure S1b, Supporting Information**). A set of SWCNT network samples was exposed to green light (525 nm, 2.36 eV) with a comparable intensity as the UV exposure. No changes in the emission spectra on any of the substrates were observed (see **Figure 6**), corroborating the role of the direct photoexcitation of $TiO_x$ or $ZnO_x$ for the introduction of luminescent oxygen defects in nanotubes.

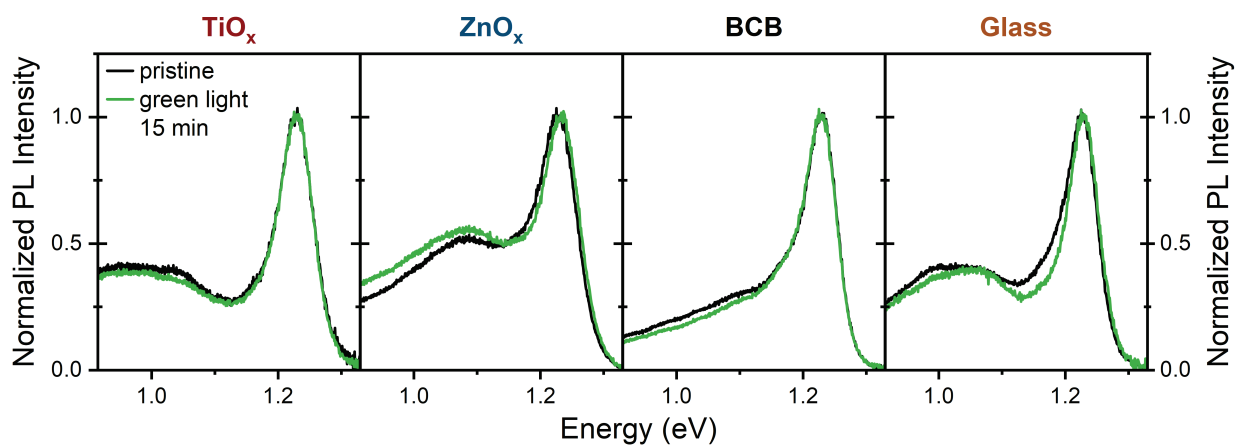

**Figure 6.** PL spectra of sparse SWCNT networks on different surfaces before and after 15 min illumination with green light (525 nm) in dry nitrogen.

To learn more about the properties of these photo-induced defect states and compare them to oxygen defects created by other methods, their averaged PL lifetimes $\tau_{avg}$ were determined by TCSPC (**Figure S7** and **Table S1, Supporting Information**). The luminescent defects in SWCNTs on TiO$_x$ showed longer PL lifetimes ($E_{11}^{*-}$ 60 ps, $E_{11}^{*}$ 26 ps) compared to those on ZnO$_x$ ($E_{11}^{*-}$ 42 ps, $E_{11}^{*}$ 25 ps), suggesting an impact of additional non-radiative decay paths such as multiphonon decay and electronic-to-vibrational energy transfer to the substrate.[58-59] The longer PL lifetimes of the $E_{11}^{*-}$ defects compared to the $E_{11}^{*}$ defects correlated with their different optical trap depths (see above) as expected.[59] Note that the lifetimes measured for these sparse networks on substrates were very short compared to defect lifetimes of functionalized nanotubes in dispersion due to increased non-radiative decay paths as observed previously. For example, Zorn *et al.* observed amplitude-averaged $E_{11}^{*}$ defect lifetimes of 222 ps in dispersion *versus* 32 to 60 ps when deposited on substrates.[29]

Furthermore, temperature-dependent PL measurements and defect/$E_{11}$ intensities (see **Figure S8a** and **S8b**, **Supporting Information**) can help to establish the thermal trap depth. The data collected from (6,5) SWCNT networks between 25 and 300 K on $TiO_x$ and $ZnO_x$ samples after 30 min UV exposure in dry nitrogen showed the expected increase of defect emission with temperature due to increasing thermal energy to overcome potential barriers for the mobile excitons around the defect sites, as reported by Kim *et al.*[60] A further increase in temperature should result in a reduction of the relative defect intensity as thermal de-trapping of excitons from the defect states starts to play a role. From this temperature range, the thermal trap depths can be extracted using van't Hoff plots of the temperature-dependent defect emission intensity relative to the $E_{11}$ emission[61] (see **Figure S8c, Supporting Information**). On $TiO_x$ the extracted thermal trap depths were 100 and 36 meV for $E_{11}*$ and $E_{11}*^-$ defects, respectively, while on $ZnO_x$ the thermal trap depth was found to be only 14 meV for $E_{11}*$. The trap depth for $E_{11}*^-$ could not be determined in the experimentally accessible temperature range (see **Table S1, Supporting Information**). Note that these thermal trap depths are comparable to those previously reported by Kim *et al.* for oxygen defects (24 meV) with similar optical trap depth.[61] However, thermal trap depths of luminescent defects also depend on defect density,[61] defect distribution and surface polarity with multiphonon decay (MPD) as a competing process.[59] A conclusive explanation for the trap depths of O-defects introduced on $ZnO_x$ and on $TiO_x$ is not possible at this point.

Finally, the selective introduction of luminescent oxygen defects in nanotube networks through UV illumination enables the creation of patterns of enhanced and red-shifted defect emission with high spatial resolution using suitable masks. Due to the necessary requirement of a dry nitrogen environment for the UV exposure, we were limited to the use of a metal shadow mask with 60 μm diameter holes. The mask was placed on top of a network of (6,5) SWCNTs on a $TiO_x$ substrate and illuminated for 30 min as schematically shown in **Figure 7a**. Subsequently, a corresponding spot of defect emission could be imaged by PL microscopy (see **Figure 7b**). The defect to $E_{11}$ emission ratio shows a very clear contrast, while the increase of the total PL signal was less well defined (see **Figure S9**, **Supporting Information**) due to the inhomogeneous distribution of SWCNTs and thus overall variations of PL intensity. Similar to the defect emission, the distribution of defects after illumination could be visualized by confocal Raman microscopy. A map of the $D/G^+$ signal ratio (see **Figure 7c**) also shows a circular pattern with an approximate diameter of 50-55 µm, representing the functionalized nanotubes. Note that the PL and Raman images were not acquired from the same spot. The slightly different shapes and diameters are most likely the result of shadow effects due to the mask thickness of 60 µm. We believe that the size of the patterned area could be reduced further by using a direct UV-laser writer or a mask aligner under inert conditions. Due to the highly local nature of the reaction, the spatial resolution should only be limited by the illumination wavelength (365 nm) and the illumination optics. It should thus be possible to reach a resolution of a few-hundred nanometers.

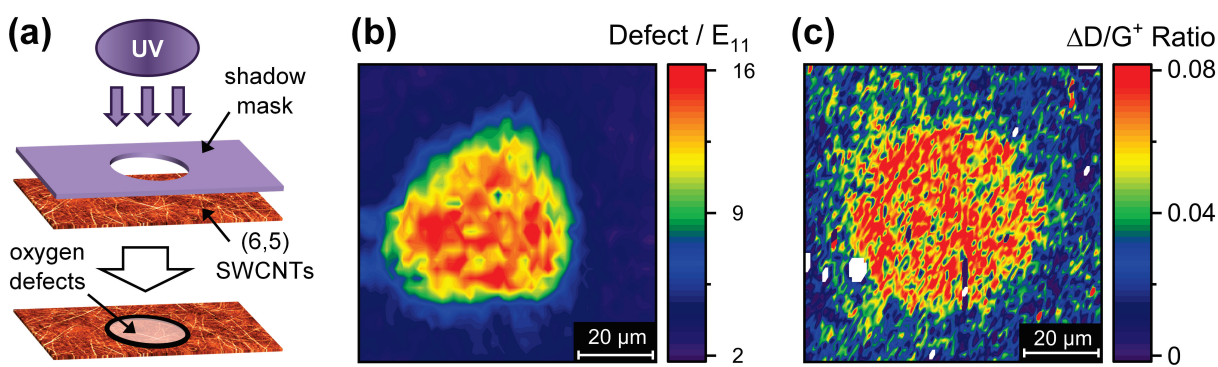

**Figure 7.** a) Schematic defect patterning process of (6,5) SWCNT networks on $TiO_x$ surface by UV exposure (30 min) through a shadow mask in dry nitrogen. Visualization of the induced luminescent defect pattern by (b) combined defect/$E_{11}$ PL ratio map and (c) Raman $D/G^+$ ratio map.

**CONCLUSIONS**

We have demonstrated the direct UV light-induced formation of luminescent oxygen defects in networks of (6,5) single-walled carbon nanotubes on photocatalytic TiO$_x$ and ZnO$_x$ surfaces under nominally inert conditions in a dry nitrogen glovebox. The reaction occurs in the solid state and does not require any solvents, additional reagents or post-treatment. We propose that water and/or oxygen adsorbed on the SWCNTs or substrate surface are oxidized/reduced by the photoexcited TiO$_x$ and ZnO$_x$ to form reactive oxygen species (ROS) that readily react with the nanotubes to form luminescent oxygen defects. Two types of emissive defects were formed, whose emission wavelengths correlate well with previously reported ether-d and epoxide-l oxygen defect configurations. The defect PL intensities varied significantly between the reactive surfaces, with defects in nanotubes on TiO$_x$ leading to overall brighter PL compared to non-illuminated SWCNTs and those irradiated on ZnO$_x$. We assume that other photoreactive oxides with different bandgaps, valence and conduction band energies (e.g., WO$_3$, In$_2$O$_3$)[62] could lead to different reactions and thus defects or defect densities depending on the preferential formation of hydroxy or superoxide anion radicals. Importantly, UV exposure in ambient air led to immediate and fast degradation of SWCNTs even on non-reactive BCB and glass substrates instead of the creation of luminescent defects, thus highlighting the importance of limiting the concentration of oxygen and water.

We applied the photo-induced formation of bright oxygen defects to create controlled emission patterns in nanotube networks by using a shadow mask during illumination. This simple approach could enable direct UV laser-writing or photolithographic patterning of oxygen defects into SWCNT networks under inert conditions and without the need for any photoresist or developer. However, our results also highlight the potential problems of photoreactive layers of TiO$_x$ or ZnO$_x$ and possibly other materials as part of optoelectronic devices with single-walled carbon nanotubes as active components. Their interactions at the interface during processing and operation (e.g., photodiodes or photovoltaic cells) may lead to unintended and undesired changes in their electronic and optical properties and thus affect short and long-term device performance.

ACKNOWLEDGEMENT

This project has received funding from the European Research Council (ERC) under the European Union's Horizon 2020 research and innovation programme (Grant Agreement No. 817494 "TRIFECTs").

# Supporting Information

Photo-Activated, Solid-State Introduction of Luminescent Oxygen Defects into Semiconducting Single-Walled Carbon Nanotubes

*Sonja Wieland, Abdurrahman Ali El Yumin, Simon Settele, Jana Zaumseil\**

Institute for Physical Chemistry, Universität Heidelberg, D-69120 Heidelberg, Germany

**Corresponding Author**

*\*zaumseil@uni-heidelberg.de*

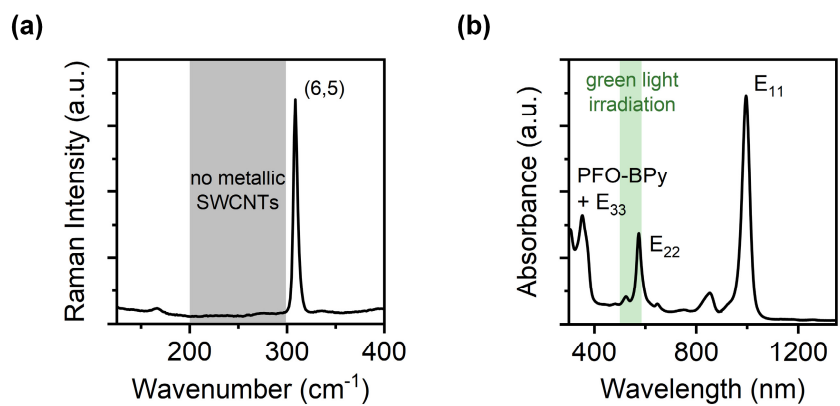

**Figure S1.** a) Resonant Raman spectrum of drop-cast (6,5) SWCNT stock dispersion (radial breathing mode range, excitation wavelength 532 nm) and b) UV-Vis-NIR absorption spectra of the dispersion in toluene after polymer-sorting and removal of excess PFO-BPy. The green light excitation wavelength range used in this work is indicated by the green box.

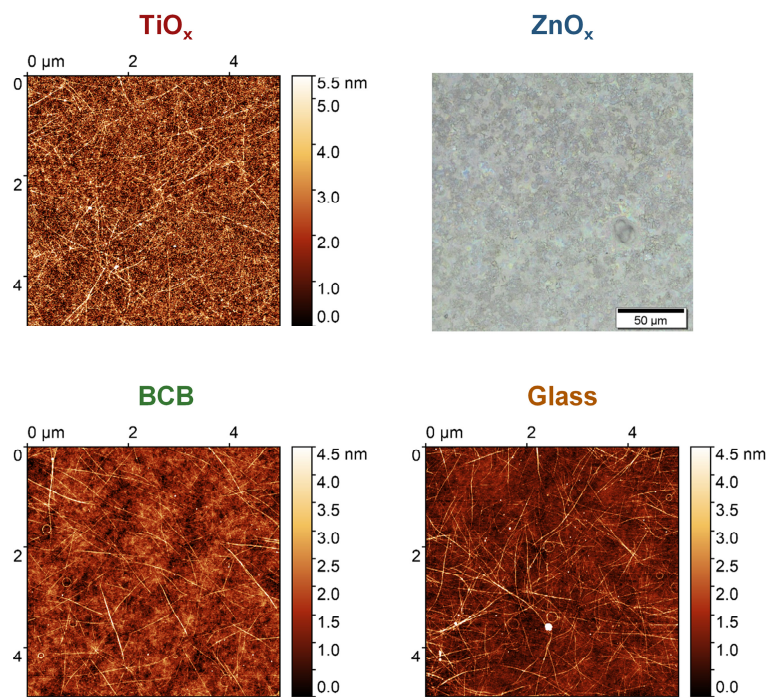

**Figure S2.** Atomic force micrographs (AFM, 5 x 5 μm²) of sparse SWCNT networks on $TiO_x$, BCB and glass, and optical micrograph of $ZnO_x$. Note that the $ZnO_x$ surface was too rough for AFM imaging.

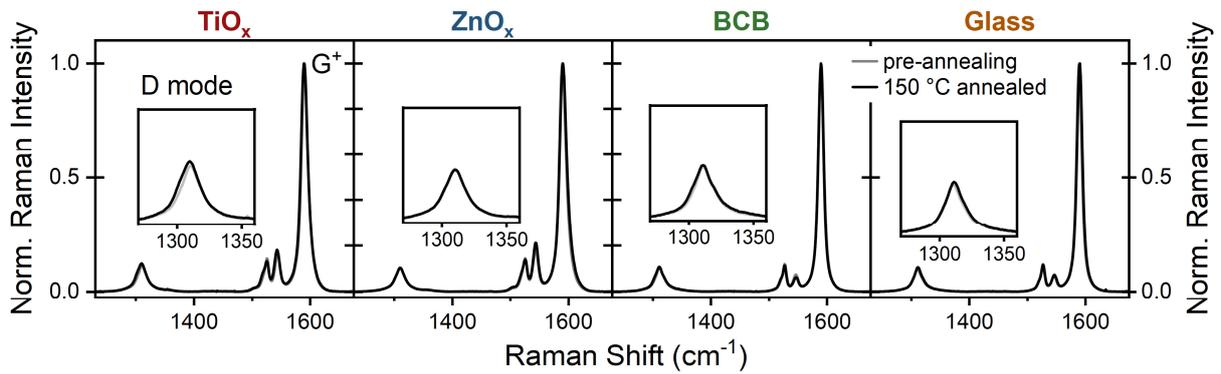

**Figure S3.** Effect of annealing at 150 °C for 30 min in dry nitrogen on Raman spectra, normalized to the $G^+$ mode. No changes in defect density are observed.

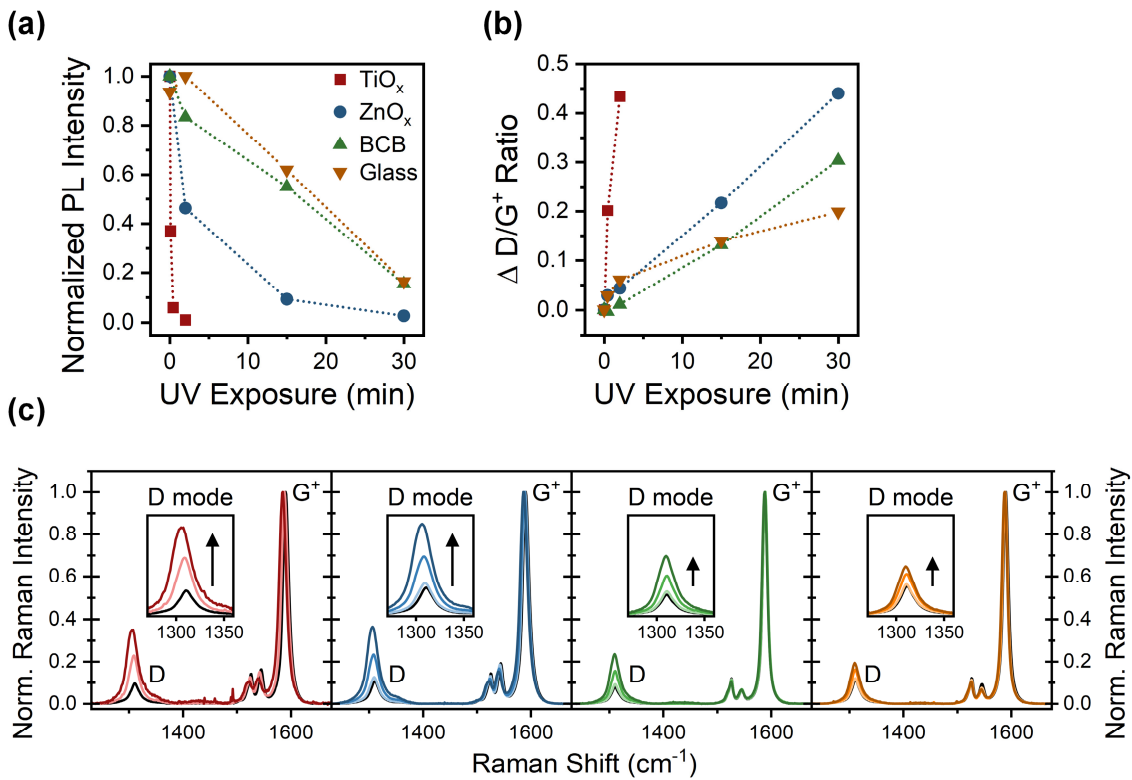

**Figure S4.** UV exposure time-dependence of a) integrated PL intensities, normalized to the respective PL intensity before illumination, and b) Raman $D/G^+$ ratios of sparse SWCNT networks on different surfaces ($TiO_x$: 0, 5, 25, 120 sec, all others: 0, 2, 15, 30 min) illuminated in air. c) Corresponding Raman spectra normalized to the $G^+$ mode.

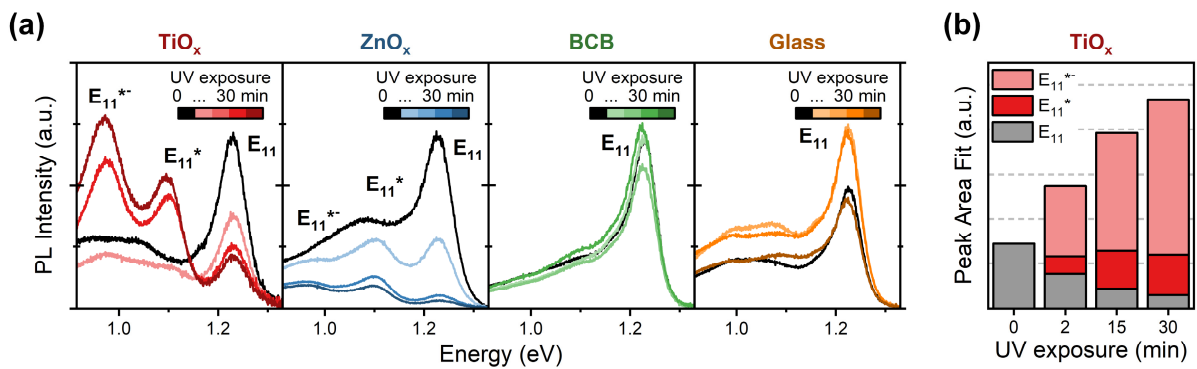

**Figure S5.** a) UV exposure time-dependent (0, 2, 15, 30 min) PL spectra of sparse SWCNT networks on different surfaces illuminated in dry nitrogen (glovebox). b) Analysis of the fitted free exciton ($E_{11}$) and defect ($E_{11}^*$, $E_{11}^{*-}$) emission intensities.

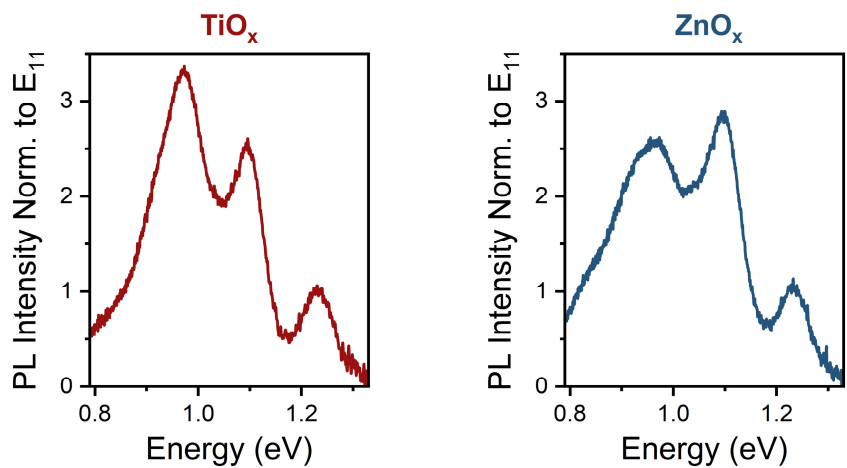

**Figure S6.** Extended-range PL spectra of SWCNTs on $TiO_x$ and $ZnO_x$ functionalized in dry nitrogen (glovebox).

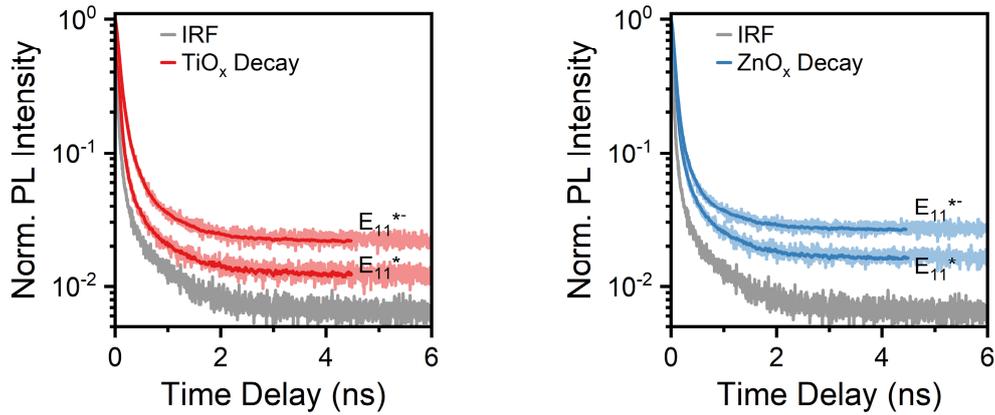

**Figure S7.** TCSPC histograms with fits of the $E_{11}^*$ and $E_{11}^{*-}$ PL decay in sparse SWCNT networks on $TiO_x$ and $ZnO_x$. The time traces were acquired at 1126 nm and 1274 nm and fitted as a biexponential decay with a re-convolution method using the fast $E_{11}$ decay of a thick SWCNT network as the instrument response function (IRF, grey).

**Table S1.** Optical ($\Delta E_{opt}$) and thermal ($\Delta E_{therm}$) trap depths, amplitude-averaged lifetimes $\tau_{avg}$, and short and long lifetime components with corresponding amplitudes (Amp) of $E_{11}^*$ and $E_{11}^{*-}$ defects on $TiO_x$ and $ZnO_x$. Lifetimes were measured at 1126 and 1274 nm (n.d. - not determined).

|  |  | $\Delta E_{opt}$ (meV) | $\Delta E_{therm}$ (meV) | $\tau_{avg}$ (ps) | $\tau_{long}$ (ps) | Amp | $\tau_{short}$ (ps) | Amp |
|---|---|---|---|---|---|---|---|---|
| $E_{11}^{*-}$ | $TiO_x$ | 268 | 36 | 60 | 194 | 0.08 | 49 | 0.92 |
|  | $ZnO_x$ | 288 | n.d. | 42 | 162 | 0.05 | 35 | 0.95 |
| $E_{11}^*$ | $TiO_x$ | 137 | 100 | 26 | 184 | 0.01 | 25 | 0.99 |
|  | $ZnO_x$ | 140 | 14 | 25 | 183 | 0.01 | 24 | 0.99 |

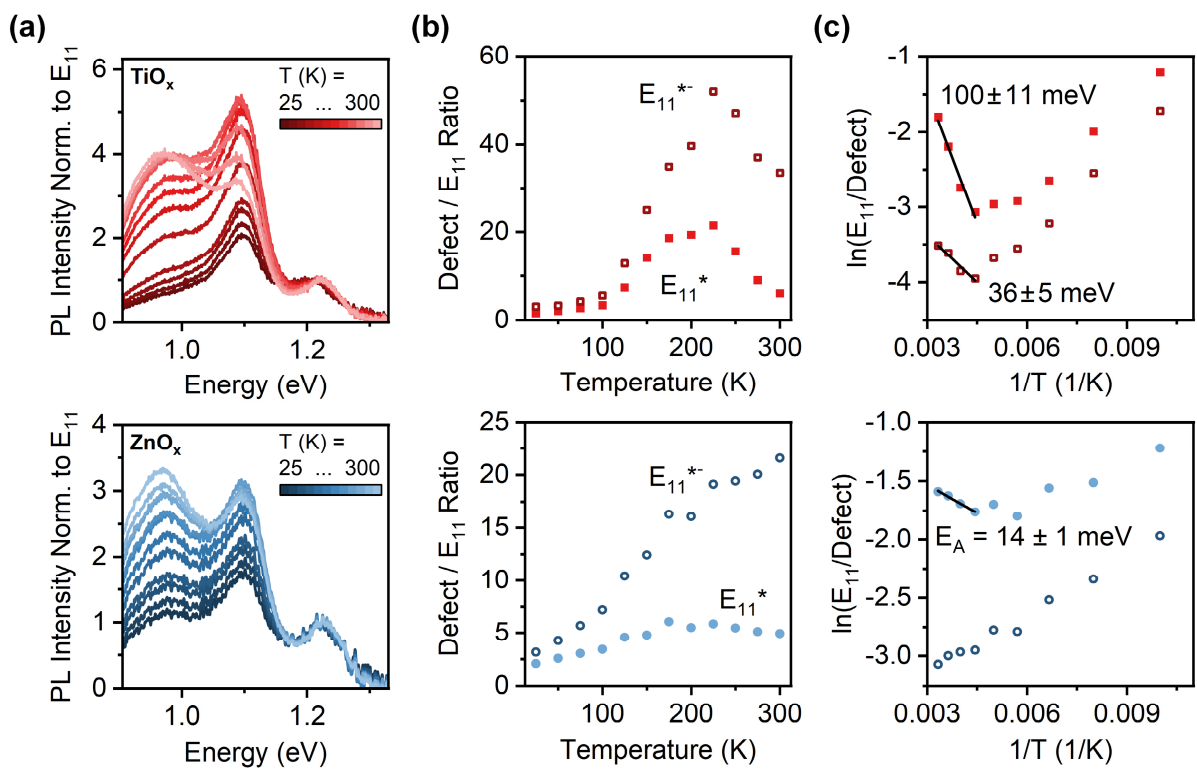

**Figure S8.** a) Temperature-dependent (25 K to 300 K in steps of 25 K) PL spectra of sparse SWCNTs on $TiO_x$ and $ZnO_x$ after defect introduction through UV exposure for 30 min in dry nitrogen (glovebox). b) Evolution of the fitted defect/$E_{11}$ PL area ratios with temperature. c) van't Hoff plots for the defect emission and linear fit to data at higher temperatures (200 to 300 K) to extract thermal de-trapping energies. Note that the extracted thermal trap depths of luminescent defects typically depend on defect density, defect distribution and surface polarity (with regard to competing multiphonon decay, MPD). A conclusive explanation for the differences between defects introduced on $ZnO_x$ and on $TiO_x$ is not possible at this point.

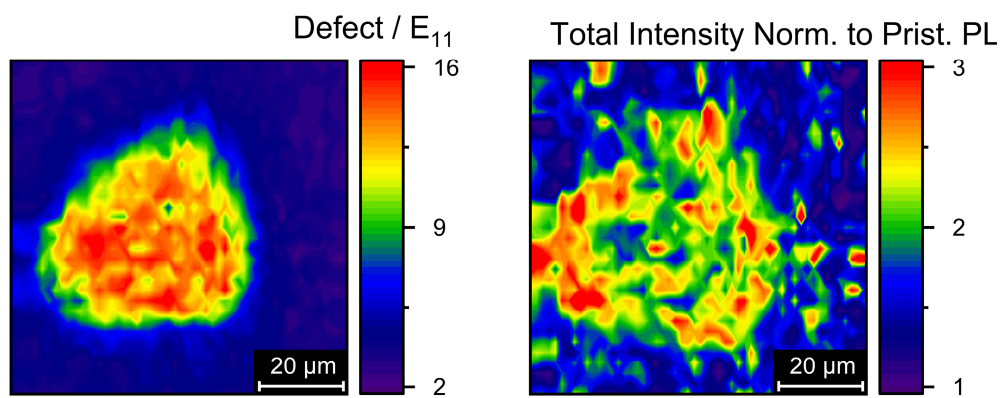

**Figure S9.** Patterned relative defect PL intensities and brightening of the overall PL. Note that variations in the SWCNT density, i.e., areas of intrinsically higher PL, result in a less defined pattern.